\documentclass[11pt,bezier,amstex]{article} 
\usepackage{mystyle-new}
\usepackage{epsfig,amsmath} 

\usepackage{color}
\usepackage{graphicx}
\definecolor{red}{rgb}{1,0,0}
\def\lesssim{\ \hbox{\raise 2pt \hbox{$<$} \kern -13pt
                     \lower 3pt \hbox{$\sim$}}\ }
\def\greatersim{\ \hbox{\raise 2pt \hbox{$>$} \kern -13pt
                     \lower 3pt \hbox{$\sim$}}\ }

\def\cascade{{\sc Cascade}}
\def\pythia{{\sc Pythia}}

\def\powheg{{\sc Powheg}}
\def\mcatnlo{{\sc Mc@nlo}}

\input epsf.tex
\def\desepsf(#1 width #2){\epsfxsize=#2 \epsfbox{#1}}
\def\kt{\ensuremath{k_t}}

\def\pt{\ensuremath{p_t}}

\newcommand{\cA}{{\cal A}}
\newcommand{\as}{\ensuremath{\alpha_\mathrm{s}}}

\def\katie{{\sc KaTie}}

\newcommand{\ccfm}{Ciafaloni:1987ur,Catani:1989yc,Catani:1989sg,Marchesini:1994wr}

\newcommand{\pd}{\partial}
\newcommand{\diff}[1]{\frac{d#1}{#1}}

\newcommand{\beq}{\begin{equation}}
\newcommand{\eeq}{\end{equation}}
\newcommand{\bea}{\begin{eqnarray}}
\newcommand{\eea}{\end{eqnarray}}
\newcommand{\beqa}{\begin{eqnarray}}
\newcommand{\eeqa}{\end{eqnarray}}

\usepackage{cite,./mcite}
\usepackage{tikz}

\usepackage{amsmath,bm}
\usepackage{lineno}

\usepackage{tikz}

\begin{document}

\begin{flushright}
DESY 17-222 \\
IFJPAN-IV-2017-28\\
version: 15 Dec 2017
\end{flushright}

\begin{center} {\sffamily\Large\bfseries
Calculations with off-shell matrix elements, TMD parton densities and TMD parton showers}
 \\ \vspace{0.5cm}

{ \Large 
Marcin~Bury$^{1}$,
Andreas ~van~Hameren$^{1}$, 
Hannes~Jung $^{1,2}$,
Krzysztof~Kutak$^{1}$, 
Sebastian~Sapeta$^{1}$,
Mirko~Serino$^{1,3}$,
}

\vspace*{0.15cm}
{\large $^1$Institute of Nuclear Physics, Polish Academy of Sciences, \\
Cracow, Poland}\\
{\large $^2$DESY, Hamburg, FRG}\\
{\large $^3$Department of Physics, Ben Gurion University of the Negev, \\ Be'er Sheva, 
Israel}\\
\end{center}
\begin{abstract}
A new calculation using off-shell matrix elements with TMD parton densities supplemented with a newly developed initial state TMD parton shower is described. The calculation is based on the \katie\ package for an automated calculation of the partonic process in high-energy factorization, making use of TMD parton densities implemented in TMDlib. The partonic events are stored in an LHE file, similar to the conventional LHE files, but now containing the transverse momenta of the initial partons. The LHE files are read in by the \cascade\  package for the full TMD parton shower, final state shower and hadronization from \pythia\ where events in HEPMC format are produced.\\
We have determined a full set of TMD parton densities and developed an initial state TMD parton shower, including all flavors following the TMD distribution.\\
As an example of application we have calculated the azimuthal de-correlation of high \pt\ dijets as measured at the LHC and found very good agreement with the measurement when including initial state TMD parton showers together with conventional final state parton showers and hadronization.
\end{abstract}

\section{Introduction} 
\label{Intro}
Measurements in today's high-energy experiments have reached a new level of precision of a few percent in experimental uncertainty. In many cases in strong interactions the theoretical predictions have larger uncertainties, mainly coming from the unknown higher order corrections which can be estimated by variation of the factorization and renormalization scales. 

While calculations in fixed order perturbation theory in Quantum Chromodynamics (QCD) even at next-to-leading (or even next-to-next-to-leading) order expansion in the strong coupling \as\ are often not sufficient, the predictions can be improved when parton showers  are included to simulate even higher order corrections, as done for example with the \powheg\ ~\mcite{Alioli:2010xa,Frixione:2007vw} or \mcatnlo\ ~\mcite{Frixione:2006gn,Frixione:2003ei,Frixione:2002bd,Frixione:2002ik} methods. However, when supplementing a calculation of collinear initial partons with parton showers, the kinematics of the hard process are changed due to the transverse momentum generated in the initial state shower \cite{Bengtsson:1986gz}. This effect can be significant even at large transverse momenta, as has been discussed and shown explicitly in \mcite{Dooling:2012uw,Hautmann:2012dw,Buckley:2016bhy}.

With the development of transverse momentum dependent (TMD) parton distributions, this problem can be overcome, since the transverse momentum of the initial partons can be obtained from the TMD parton distributions.  The great advantage of using TMD parton densities is that a  parton shower will not change the kinematics of the matrix element process, in contrast to the conventional approach of collinear hard process calculations supplemented with parton showers, and that the main parameters of the TMD parton shower are fixed with the determination of the TMD.

Already some time ago a TMD parton shower has been developed for the case of initial state gluons within the frame of the CCFM evolution equation \cite{Marchesini:1994wr,Catani:1989sg,Catani:1989yc,Ciafaloni:1987ur} and implemented in the \cascade\ package \mcite{Jung:2010si,Jung:2001hx,Jung:2000hk,Marchesini:1992jw,Marchesini:1990zy}.  However, TMD parton densities defined over a large range in $x$, \kt\  and scale $\mu$ for all different flavors including quarks and gluons were not available until recently. In  \cite{Hautmann:2017fcj,Hautmann:2017xtx} a new method for determination of TMD parton densities is described, another method to obtain TMD parton densities from collinear parton densities has been proposed in \cite{Martin:2009ii}, which we apply in the present study. In order to fully account for the potential of a TMD parton shower, the initial state kinematics for the hard process calculation should include the transverse momenta. With the development of an automated calculation of multi-leg matrix elements with off-shell initial states \cite{vanHameren:2016kkz} the full potential of TMD parton densities and parton showers can be explored.

In this article we will describe how the TMD parton densities can be obtained from the KMRW approach \cite{Martin:2009ii} and how they can be used in calculations using off-shell matrix elements obtained from \katie\  \cite{vanHameren:2016kkz}. We then describe how this matrix element calculation is supplemented with a newly developed TMD parton shower, which makes use of the TMD parton densities without changing the kinematics of the matrix element process. We illustrate the advantage of using TMD densities with off-shell matrix element calculations in an application to azimuthal de-correlations of high \pt\  dijet measurements at the LHC.

In section~\ref{TMDme} we briefly describe the main features of the automated calculation of off-shell matrix elements with \katie\ and section~\ref{TMDpdf} describes the procedure to obtain the TMD parton densities with the KMRW method. In section~\ref{TMDshower} we describe a new development of the TMD parton shower which can be combined with the matrix element calculation via LHE files, similar to what is being used in standard methods. In section~\ref{Results} we present a case-study of azimuthal correlations of dijets at large transverse momenta as obtained at the LHC.

\section{Off-shell matrix element calculation and partonic cross section}
\label{TMDme}
\katie\ is a parton-level event generator for arbitrary processes within the Standard Model, with the special feature that it can generate events with space-like initial-state momenta that have non-vanishing transverse components.
It produces weighted parton-level event files in the Les Houches format  \cite{Alwall:2006yp}, or in a custom format.
For the latter, \katie\ also provides the tools to produce distributions for arbitrary observables.
It relies on LHAPDF~\cite{Buckley:2014ana} for collinear PDFs and the running coupling constant, and on TMDlib~\cite{Hautmann:2014kza} for transverse momentum dependent PDFs.
Alternatively, the latter can be provided as hyper-rectangular grids which {\sc KaTie} itself interpolates.
The hard matrix elements are calculated as the summed squares of helicity amplitudes, defined following the approach of~\cite{vanHameren:2012if,vanHameren:2013csa} which guarantees gauge invariance.
The amplitudes are calculated numerically with recursive methods~\cite{Berends:1987me,Caravaglios:1995cd} which keep the computational complexity under control, even for larger final-state multiplicities.

A project is defined in a single user-defined input file, containing all the information about the desired center-of-mass energy, inclusive phase space cuts, and values of model parameters like particles masses and widths.
If the user wants to apply TMDPDFs that are not included in TMDlib, this file must also include the paths to the files containing the hyper-rectangular grids.
Finally, \katie\ does not generate a list of partonic sub-processes itself, and the user must provide this list in the same input file.

Event generation happens in two stages.
During the fist stage, the phase space sampler is optimized for each sub-process separately.
This stage is very cheap in terms of CPU time compared to the second stage during which the actual event files are generated.
This stage can trivially be parallelized by running several instances of the executable with different seeds for the random number generator.

\section{TMD parton density functions}
\label{TMDpdf}
The complete set of transverse momentum dependent PDFs consistent with the matrix elements that we use can be obtained by applying Lipatov's effective action 
approach combined with the Curci-Furmanski-Petronzio method, which allows to formally define new splitting functions. 
The construction of a new set of evolution equations and the corresponding parton densities is still to be achieved. 
Only recently all the real contributions to the TMD splitting functions have been obtained \cite{Hentschinski:2017ayz}. 
At present, we obtain TMD parton densities from collinear parton densities by the application of the KMRW procedure~\cite{Martin:2009ii}. 
In this method the \kt-dependent distributions are calculated from the DGLAP equation by taking into account only the contribution corresponding to a single real emission. 
The virtual contributions between the scales \kt\ and $\mu$ are resummed into a Sudakov factor, which describes the probability that there are no emissions. 

The precise expressions for the TMD distributions read
\beq
\mathcal{A}_i(x, \kt^2, \mu^2) = \frac{\pd}{\pd \kt^2}\left[ xf_i(x, \kt^2) \, \Delta_i(\kt^2, \mu^2) \right]
\eeq
with the Sudakov factors for quarks
\beq
  \Delta_q(\kt^2,\mu^2) = \exp\left(-\int_{\kt^2}^{\mu^2}\!\diff{\kappa_t^2}\,\frac{\alpha_S(\kappa_t^2)}{2\pi}\,\int_0^1\! d{\zeta }\,P_{qq}(\zeta )\Theta(1-z_M-\zeta)\right)
  \label{eq:Tq}
\eeq
and for gluons
\beq
  \Delta_g(\kt^2,\mu^2) = \exp\left(-\int_{\kt^2}^{\mu^2}\!\diff{\kappa_t^2}\,\frac{\alpha_S(\kappa_t^2)}{2\pi}\, \int^1_0\!d{\zeta }\;[~\zeta \,P_{gg}(\zeta )\Theta(1-z_M-\zeta)\Theta(\zeta-z_M) + n_F P_{qg}(\zeta)~]\right).
  \label{eq:Tg}
\eeq
Here, $n_F$ is the active number of quark--antiquark flavours into which the gluon may split, and we set $n_F=5$. The infrared cutoff $z_M~\equiv~ \frac{\kt}{\mu+\kt}$ arises because of the singular behaviour of the splitting functions $P_{qq}(z)$ and $P_{gg}(z)$ at $z=1$, which correspond to soft gluon emission.

\begin{figure}[bhtb]
\begin{center}
\includegraphics[width=0.48\textwidth]{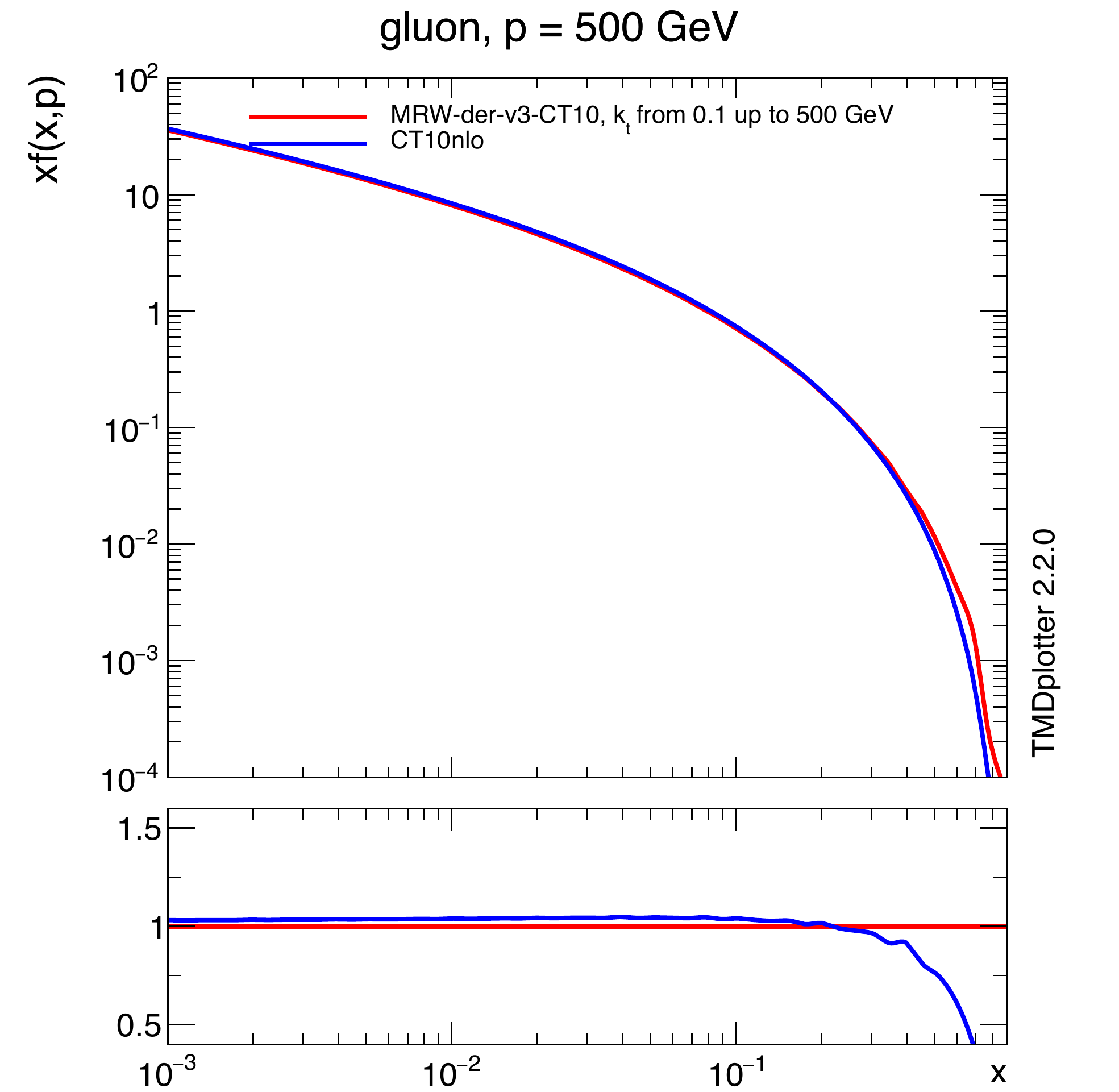} \hskip 0.2cm
\includegraphics[width=0.48\textwidth]{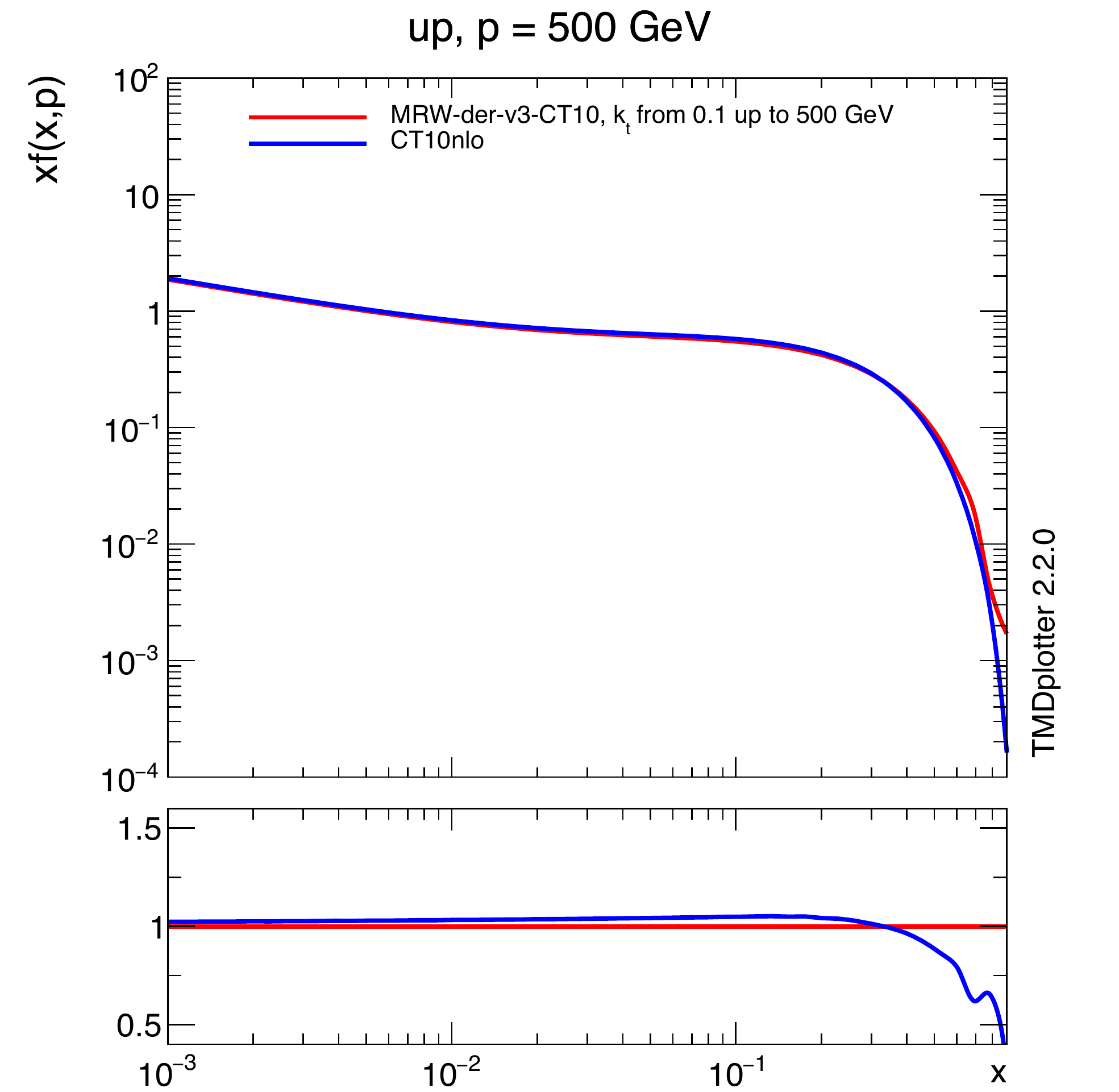} 
\caption{Comparison between the integrated TMD using the method of Ref.~\protect\cite{Martin:2009ii}  and the underlying collinear CT10nlo gluon PDFs~\protect\cite{Lai:2010vv} at a scale $\mu=500$~GeV for gluons (left) and u-quarks (right).}
\label{integratedPDFs}
\end{center}
\end{figure}

The TMDs are defined only for $\kt > \mu_0$, where $\mu_0 \sim 1$ GeV is the minimum scale for the the integrated (collinear) PDFs. 
In order to extend the TMD to the region  $\kt < \mu_0$, we tested three methods. 
One is to set the TMD proportional to \kt , the second is to freeze the TMD at $\kt = \mu_0$ and the third is taken from Ref.~\cite{Martin:2009ii} and is used here:%
\beq \label{eq:lowkt}
\mathcal{A}_i(x, \kt^2, \mu^2) = \frac{1}{\mu_0^2}\,xf_i(x, \mu_0^2)\,\Delta_i(\mu_0^2, \mu^2).
\eeq

The TMDs used here (MRW-CT10nlo) are based on the CT10nlo collinear PDF set~\cite{Lai:2010vv}  including the appropriate running coupling \as . In fig.~\ref{integratedPDFs} we show a comparison of the original CT10 parton density with the TMDs constructed here integrated over \kt\  up to the scale $\mu$ using the TMDplotter tool \cite{Hautmann:2014kza,TMDplotter2}. We observe reasonable agreement, except at large $x$,  where the integration limits in the Sudakov form factor play a role. The large $x$ region is, however, not relevant for the processes studied here.

\begin{figure}[htb]
\begin{center}
\includegraphics[width=0.48\textwidth]{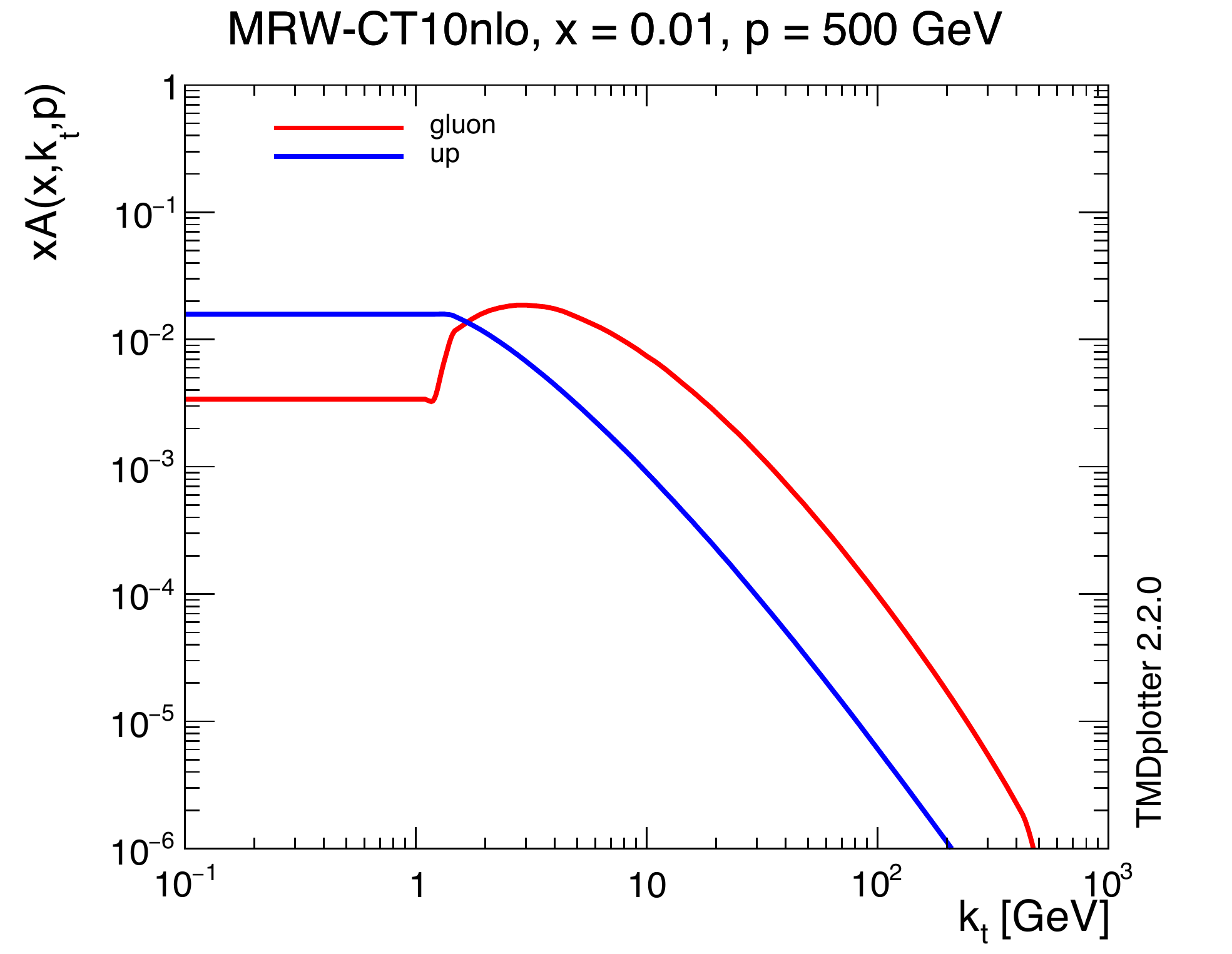} \hskip 0.2cm
\includegraphics[width=0.48\textwidth]{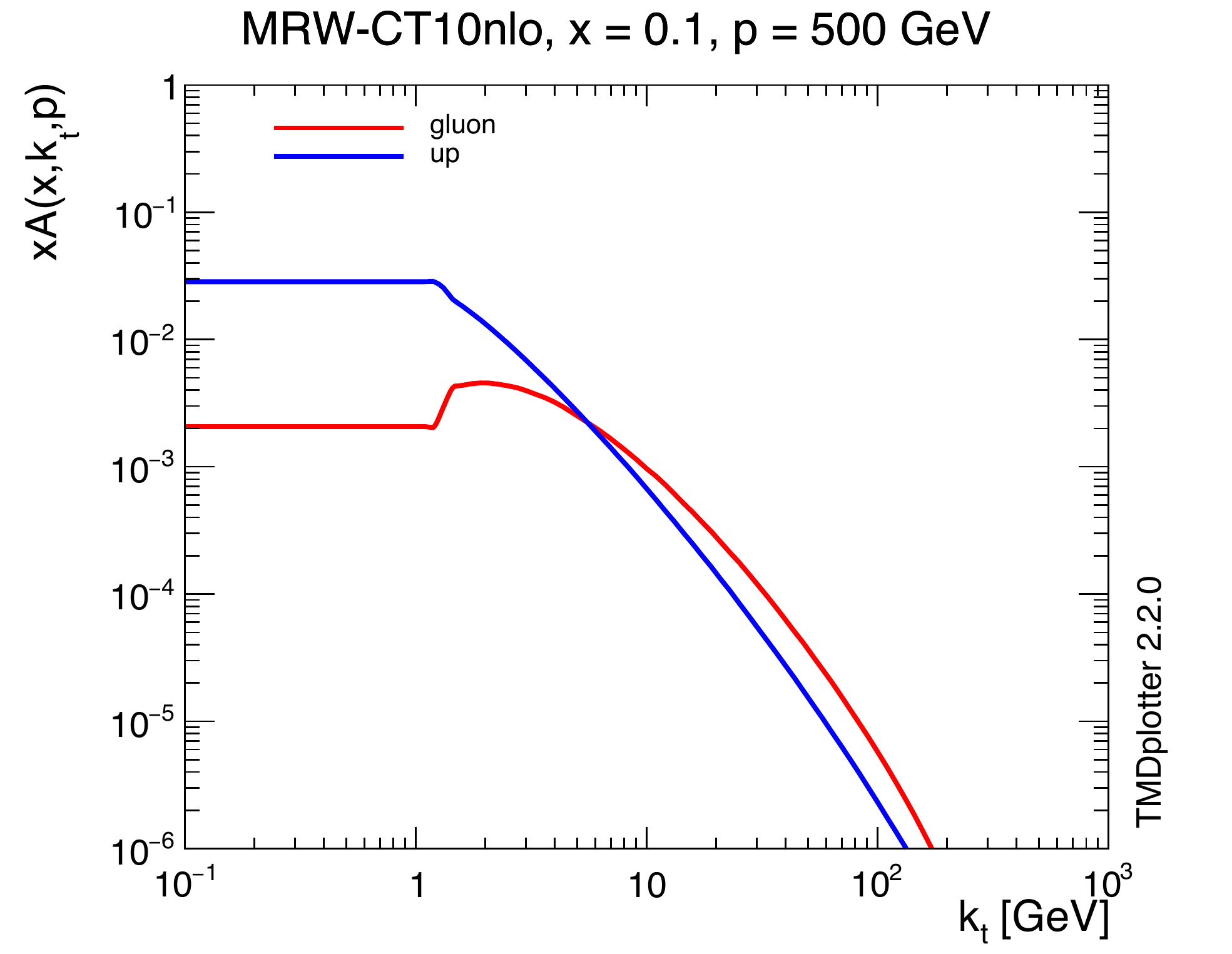} 
\caption{Transverse momentum distribution of the TMD at a scale $\mu=500$~GeV for gluons and u-quarks at $x=0.01$ (left) and $x=0.1$ (right).}
\label{TMDs}
\end{center}
\end{figure}

In fig.~\ref{TMDs} we show the \kt\ dependence of the TMD at a scale $\mu=500$~GeV for  different values of $x$. One can clearly see the treatment of the non-perturbative region of $\kt < 1$~GeV. The discontinuity at small \kt\ comes from the matching procedure in eq.(\ref{eq:lowkt}).

\section{Initial State Parton Shower based on TMDs}
\label{TMDshower}
The parton shower, which is described here, follows consistently the parton evolution of the TMDs. 
By this we mean that the splitting functions $P_{ab}$, the order in \as , the scale in the calculation of \as\, as well as the kinematic restrictions applied are identical in both the parton shower and the evolution of the parton densities.

A backward evolution method, as now common in Monte Carlo event generators, is applied for the initial state parton shower, evolving from the large scale of the matrix-element process backwards down to the scale of the incoming hadron. However, in contrast to the conventional parton shower, which generates a transverse momentum of the initial state partons during the backward evolution, the transverse momentum of the initial partons of the hard scattering process is fixed by the TMD and the parton shower does not change the kinematics.
The transverse momenta during the cascade follow the behavior of the TMD.  The hard scattering process is obtained directly using off-shell matrix element calculations as described in section~\ref{TMDme}. The partonic configuration is stored in the form of an LHE (Les Houches Event) text file, but now including the transverse momenta of the incoming partons. This LHE files are input to the shower and hadronization interface of \cascade \cite{Jung:2010si,Jung:2001hx} (new version \verb+2.4.X+) for the TMD shower where events in HEPMC~\cite{Dobbs:2001ck} format are produced. 

The backward evolution of the initial state parton shower follows very closely the description in \mcite{Bengtsson:1986gz,Jung:2010si,Jung:2001hx,Jung:2000hk}. The evolution scale $\mu$ is selected from the hard scattering process, with $\mu^2 = \hat{p}_T^2$  or $\mu^2 = Q_t^2 +\hat{s}$ for an evolution in virtuality or angular ordering, with $\hat{p}_T$ being the transverse momentum of the hard process, $Q_t$ being the vectorial sum of the initial state transverse momenta and $s$ being the invariant mass of the subprocess.  

\begin{figure}[htb]
\begin{center} 
\includegraphics[width=0.25\textwidth]{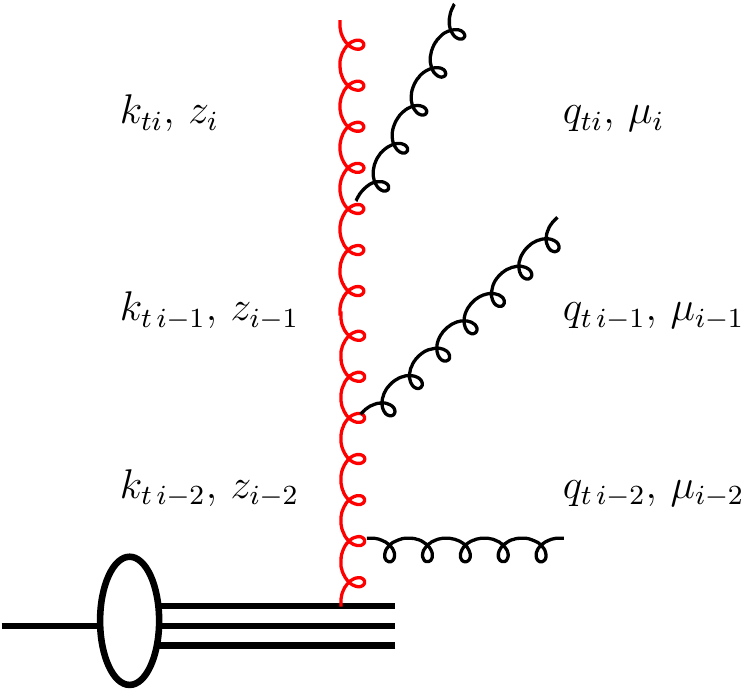} \hskip 2cm
\includegraphics[width=0.2\textwidth]{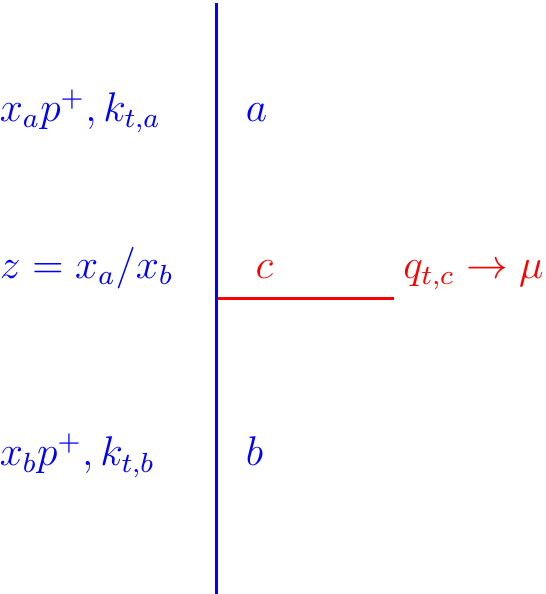} 
  \caption{Left: Schematic view of a parton branching process. Right: Branching process $ b \to a + c$.}
\label{parton-branching}
\end{center}
\end{figure}

Starting with the hard scale $\mu=\mu_{i}$, the  parton shower algorithm searches for the next scale $\mu_{i-1}$ at which a resolvable branching occurs. This scale $\mu_{i-1}$ is selected from the Sudakov form factor $\Delta_S$ making use of the TMD densities $\cA_a(x',\kt',\mu')$ which depend on the longitudinal momentum fraction $x'=x z$ of parton $a$, its transverse momentum $\kt'$ probed at a scale $\mu'$ (see also \cite{Jung:2010si}). The Sudakov form factor $\Delta_S$ for the backward evolution is given by (see fig.~\ref{parton-branching} left):
\begin{equation}
\Delta_S(x,\mu_{i},\mu_{i-1}) = \exp\left[ - \int_{\mu_{i-1}}^{\mu_{i}} \frac{d \mu'}{\mu'} \frac{\as({\tilde \mu'})}{2 \pi} \sum_a \int dz P_{a\to bc}(z) \frac{x'\cA_a(x',\kt',\mu')}{x\cA_b(x,\kt,\mu')} \right]
\label{sudakov}
\end{equation}
which describes the probability that parton $b$ remains at $x$ with transverse momentum $\kt$ when evolving from $\mu_i$ to $\mu_{i-1} < \mu$. Please note, that the argument in \as\ is $\tilde \mu'$ and depends on the ordering condition as discussed later.
\footnote{In equation eq.(\ref{sudakov}) ordering in $\mu$ is assumed, if  angular ordering, as in CCFM~\cite{\ccfm}, is applied then the ratio of parton densities would change to $ \frac{x'\cA_a(x',\kt',\mu'/z)}{x\cA_b(x,\kt,\mu')}$ as discussed in \cite{Jung:2010si}.}

In the parton shower language, the selection of the next branching comes from solving the Sudakov form factor eq.(\ref{sudakov}) for $\mu_{i-1}$.
However, to solve the integrals in eq.(\ref{sudakov}) numerically for every branching would be too time consuming, instead the veto-algorithm \cite{Bengtsson:1986gz,Platzer:2011dq} is applied. The selection of $\mu_{i-1}$ and the branching splitting $z_{i-1}$ follows the standard methods \cite{Bengtsson:1986gz}. 

The splitting function $P_{ab}$ as well as the argument $\tilde \mu$ in the calculation of \as\ is chosen exactly as used in the evolution of the parton density. In a parton shower one treats ``resolvable'' branchings, defined via a cut in $z < z_M$ in the splitting function  (see eq.(\ref{eq:Tg})) to avoid the singular behavior of the terms $\frac{1}{1-z}$, and branchings with $z> z_M$ are regarded as ``non-resolvable'' and are treated similarly as virtual corrections: they are included in the Sudakov form factor $\Delta_S $. 

The longitudinal momentum fraction $x_{i-1}= \frac{x_i}{z_{i-1}}$ is calculated by generating $z_{i-1}$ according to the splitting function. With $z_{i-1}$ and $\mu_{i-1}$ all variables needed for a collinear parton shower are obtained. 

The calculation of the  transverse momentum $\kt$ is sketched in fig.~\ref{parton-branching} right.
The transverse momentum $q_{t\,i}$ can be  obtained by giving a physical interpretation to the evolution scale $\mu_i$ (see fig.~\ref{parton-branching} right), and $q_{t\,i}$ can be calculated in case of angular ordering ($\mu$ is associated with the angle of the emission)  in terms of the angle $\Theta$ of the emitted parton wrt the beam directions $q_{t,c} = (1-z) E_{b} \sin \Theta$:
\begin{equation}
  \label{ang-ordering}
 {\bf q}_{t,i}^2  =  (1-z)^2 \mu_i^ 2  \;\; .
\end{equation}

Once the transverse momentum of the emitted parton $q_t$ is known, the transverse momentum of the propagating parton can be calculated from
\begin{equation}
{\bf k}_{t\,i-1} = {\bf k}_{t\,i} + {\bf q}_{t\, i-1}
\end{equation}
with a uniformly distributed azimuthal angle $\phi$ is assumed for the vector components of $\bf k$ and $\bf q$.

The whole procedure is iterated until one reaches a scale $\mu_{i-1} < q_0$ with $q_0$ being a cut-off parameter, which can be chosen to be the starting evolution scale of the TMD. However, it turns out that during the backward evolution the transverse momentum $k_t$ can reach large values, even for small scales $\mu_{i-1}$, because of the random $\phi$ distribution. On average the transverse momentum decreases, and it is of advantage to continue the parton shower evolution to a scale $q_0 \sim \Lambda_{qcd} \sim 0.3$~GeV, to allow enough emissions to share the transverse momenta generated.

\section{Predictions for high \pt\ dijets in pp at the LHC}
\label{Results}
We show predictions obtained with off-shell matrix elements of $2\to2$ QCD processes using the TMDs obtained in sec.~\ref{TMDpdf}.
The results of the parton level calculation are fed via LHE files to the shower and hadronization interface of \cascade \cite{Jung:2010si,Jung:2001hx} (new version \verb+2.4.X+) for the TMD shower where events in HEPMC format are stored for further processing as via Rivet \cite{Buckley:2010ar}. 

\begin{figure}[htb]
\begin{center}
\includegraphics[width=0.45\textwidth]{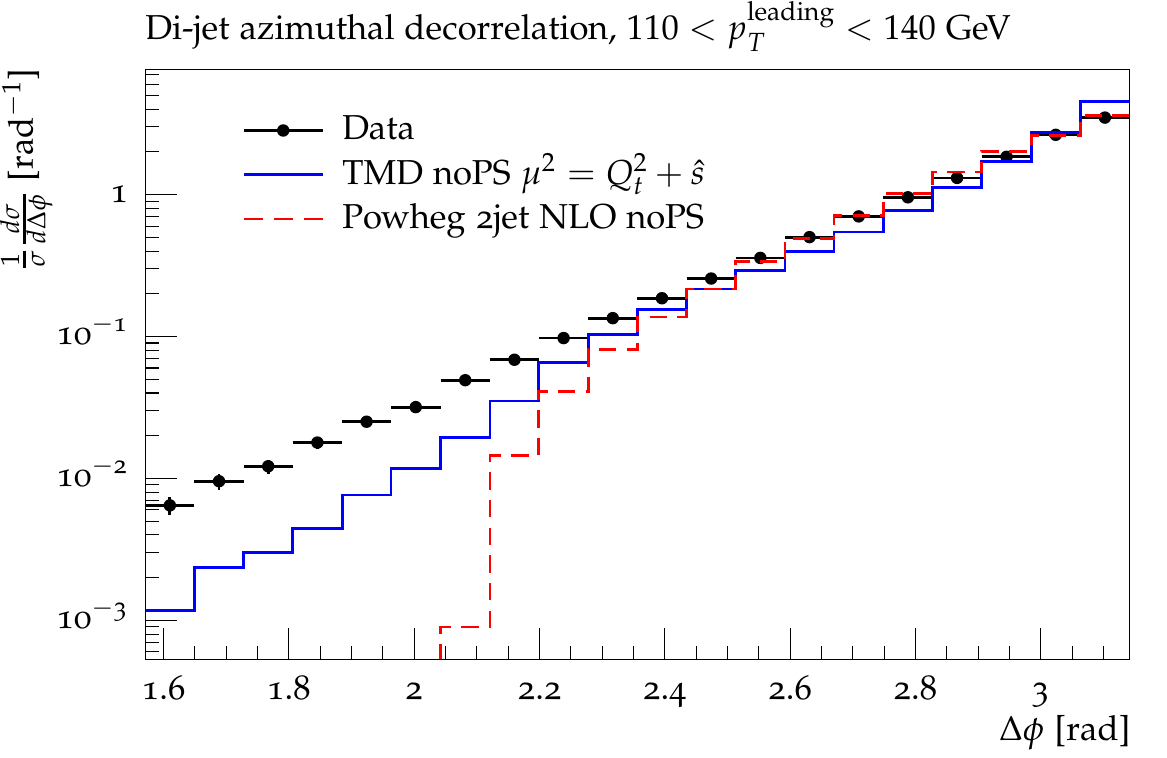} \hskip 1cm
\includegraphics[width=0.45\textwidth]{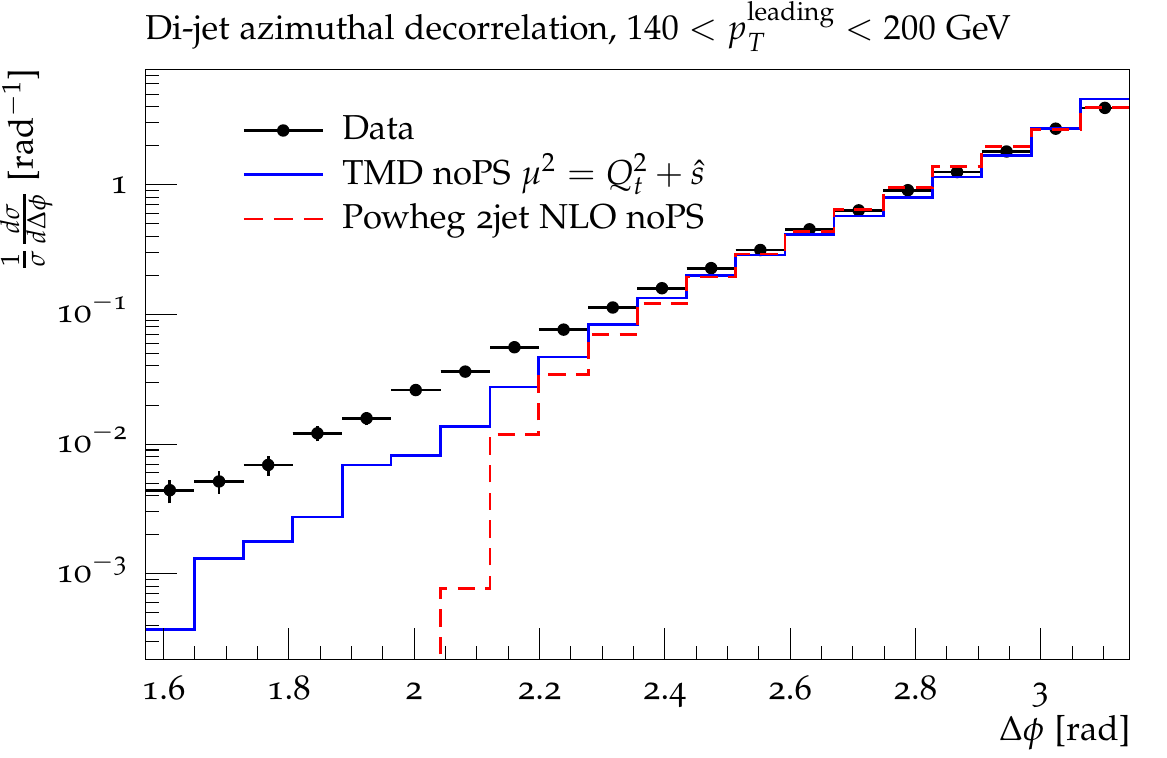} 
\caption{$\Delta \phi$ distribution for high \pt\ dijet production \protect \cite{Khachatryan:2011zj}. The solid (blue) histogram shows the prediction using off-shell $2\to2$ matrix elements with TMD parton densities, the dashed (red) line is a 3-parton configuration obtained with \powheg . Both predictions are without parton shower and hadronization. }
\label{dijets_PH}
\end{center}
\end{figure}

First we show parton level results of azimuthal de-correlations of high \pt\ dijet production at the LHC at $\sqrt{s}=7$~TeV \cite{Khachatryan:2011zj}. In fig.~\ref{dijets_PH} we compare predictions obtained from our calculation (without parton shower)  with the one from \powheg\ dijet (without parton shower). One can observe reasonable agreement between both parton level calculations at high $\Delta \phi$. The \powheg\ prediction shows a sharp drop at $\Delta \phi = 2\pi /3$, which is the kinematic limit for a 3 parton configuration. The prediction using TMDs shows a smooth distribution to smaller values of $\Delta \phi$ which is typical for a configuration where more partons are radiated in the initial state. The distribution of our prediction depends entirely on the shape of the TMD.
Thus, with a precise determination of the TMD, we expect the $\Delta \phi$ distribution to be well described, without any tuning and without any adjustment of additional parameters.

\subsection{Predictions including TMD parton showers}

In fig.~\ref{dijets_PH} we have shown the advantage in using TMD parton densities compared to a fixed order collinear calculation: due to the resummation of multiple parton emissions in the TMD parton density, the phase space for multi-jet production is covered, as seen in the tail to small $\Delta \phi$. Of course, the experimental measurement is different from a purely 2-parton final state, even using TMDs, since the jet clustering is based on multiple partons (hadrons).
\begin{figure}[htb]
\begin{center}
\includegraphics[width=0.45\textwidth]{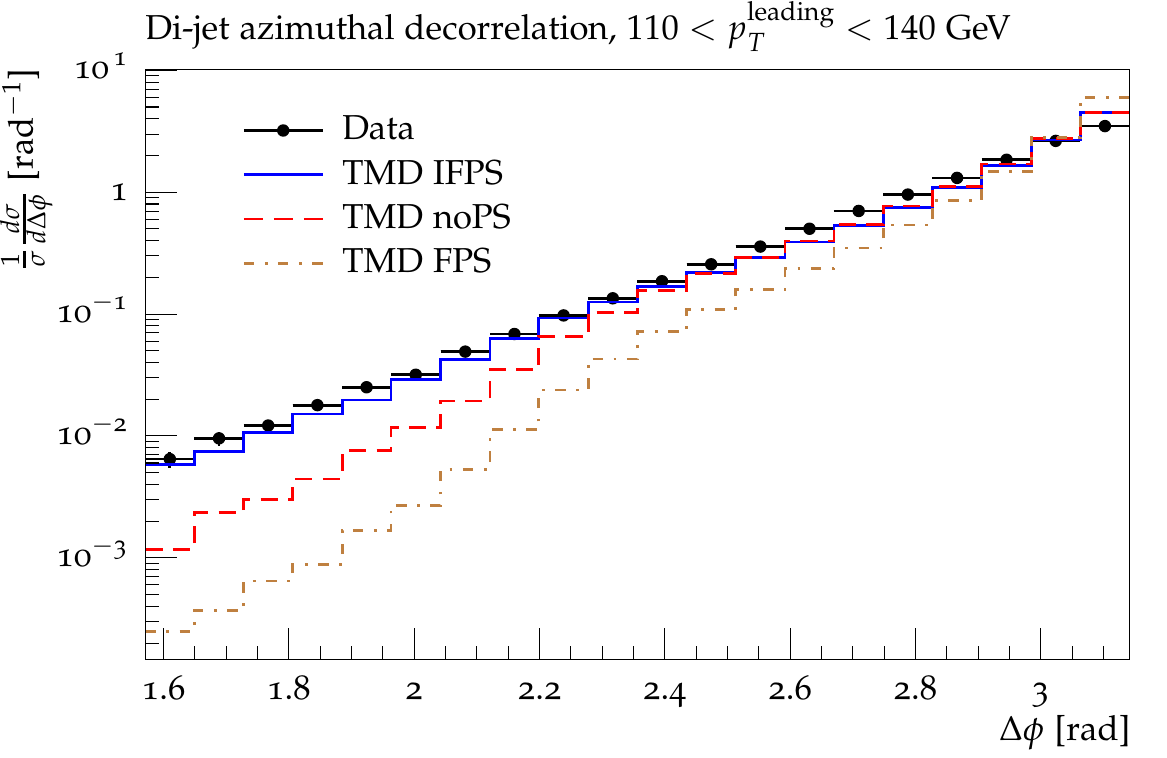} \hskip 1cm
\includegraphics[width=0.45\textwidth]{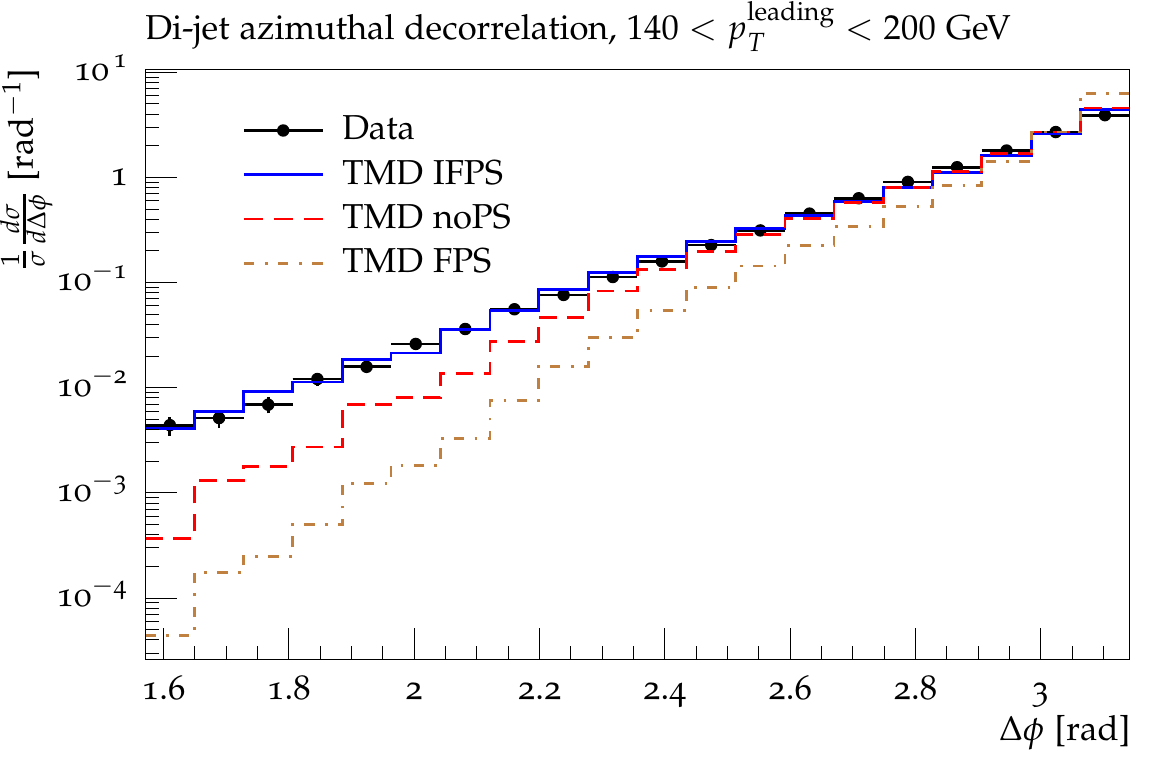} 
\caption{$\Delta \phi$ distribution of high \pt dijet events for different regions of $p_t^{leading}$: without parton shower (noPS, dashed red line), with final state parton shower (FPS, dashed-dotted brown line), with initial TMD shower and final state parton shower (IFPS, blue solid line). The factorization scale $\mu^2=  Q_t^2 + \hat{s}$ was chosen.}
\label{dijets_shower}
\end{center}
\end{figure}
In fig.~\ref{dijets_shower} we show a comparison of the prediction using TMDs with and without initial state TMD parton showering and including final state parton shower and hadronization (taken from \pythia~\cite{Sjostrand:2006za}), with a final state parton shower scale of $\mu_{fps} = 2 \hat{p}_t $  being the average  transverse momentum of the outgoing matrix element partons. While even without parton shower a tail towards small $\Delta \phi$ is observed, the simulation of the parton shower, both initial TMD and final state parton shower contributes to the shape of the distribution and  brings it close to the measurement.

\begin{figure}[htb]
\begin{center}
\includegraphics[width=0.45\textwidth]{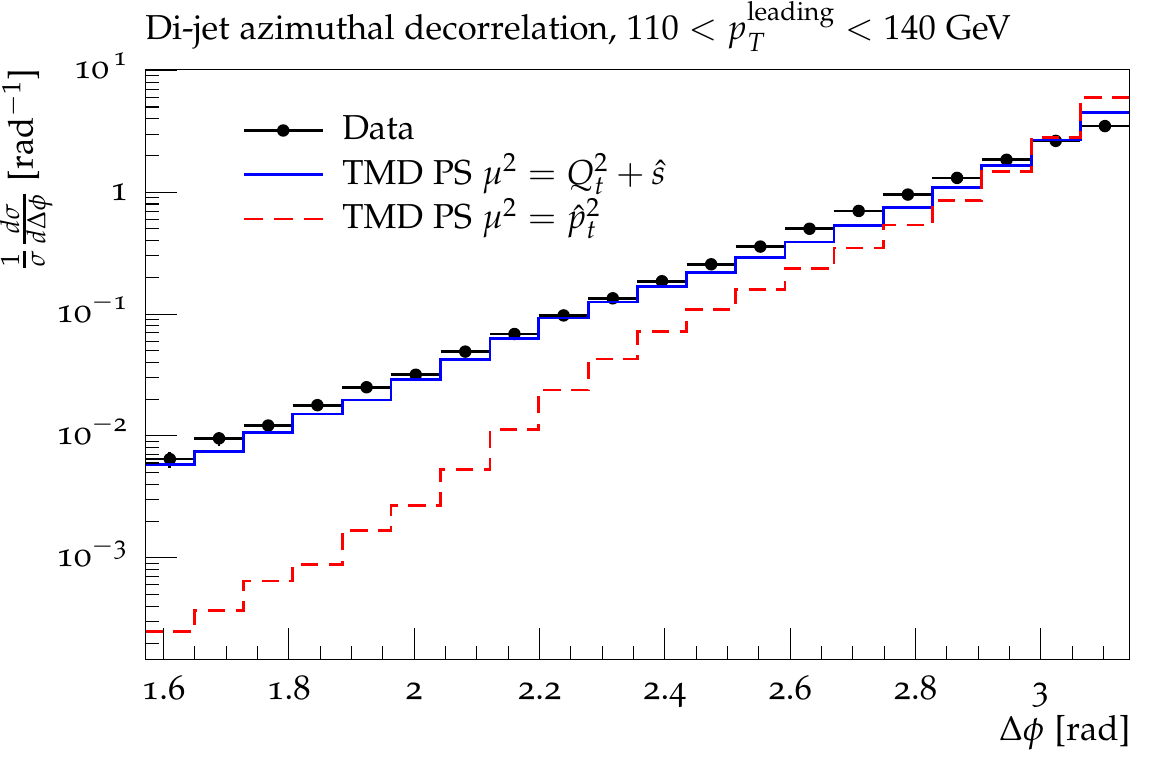} \hskip 1cm
\includegraphics[width=0.45\textwidth]{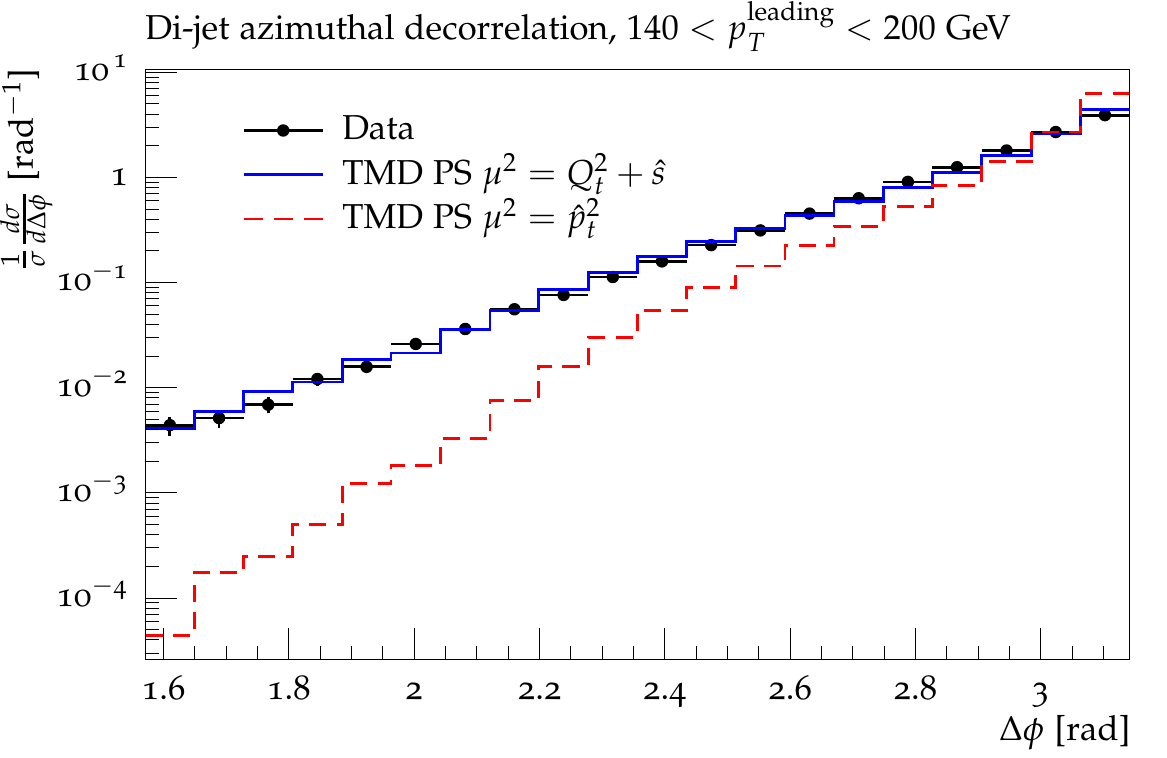}  
\includegraphics[width=0.45\textwidth]{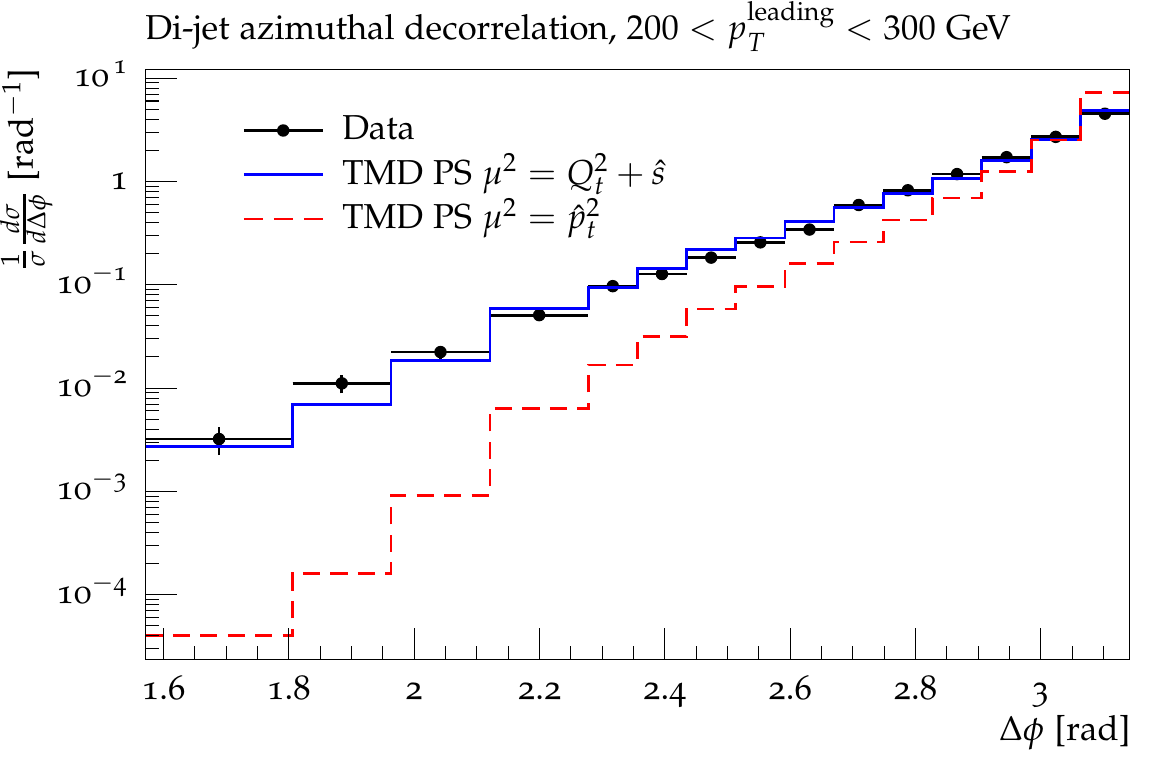} \hskip 1cm
\includegraphics[width=0.45\textwidth]{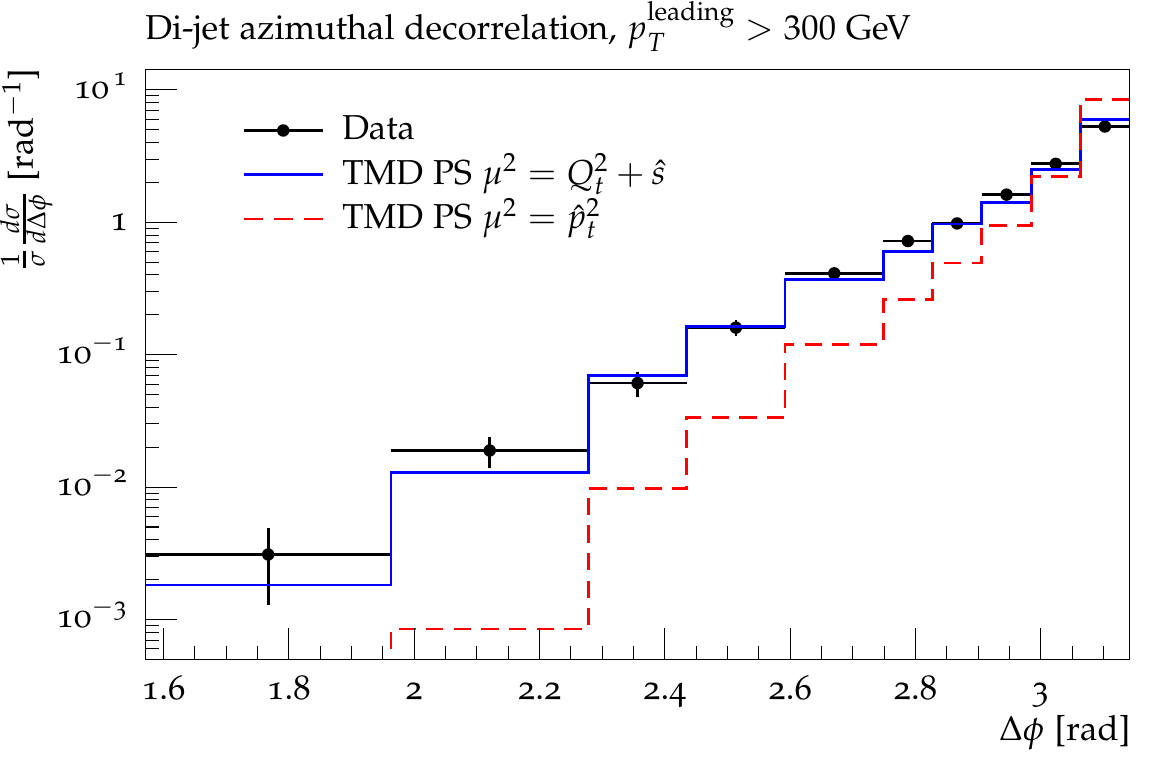} 
\caption{$\Delta \phi$ distribution as measured by \protect\cite{Khachatryan:2011zj} for different regions of $\pt^{leading}$. The data are compared with predictions using off-shell $2\to2$ matrix elements with TMD parton densities, an initial state TMD parton shower, conventional final state parton shower and hadronization. Shown are predictions for two different choices of the factorization scale, as discussed in the text.}
\label{dijets_shower_data}
\end{center}
\end{figure}

In fig.~\ref{dijets_shower_data} we show predictions for the azimuthal de-correlation $\Delta \phi$ for high \pt\ dijets for different regions of $\pt^{leading}$ using TMD parton densities with off-shell matrix elements, parton shower and hadronization in comparison with measurements at $\sqrt{s} = 7$ TeV in pp collisions at the LHC \cite{Khachatryan:2011zj}. We show predictions for two different factorization scales:
$\mu^2 = Q_t^2 + \hat{s}$, where $Q_t$ is the vectorial sum of the initial state transverse momenta and $\sqrt{\hat{s}}$ is the invariant mass of the partonic subsystem and $\mu^2 =\hat{p_t}^2$.
The first scale choice is motivated by angular ordering (see Ref. \cite{Jung:2003wu}), the second one is the conventional scale choice. The scale choice motivated from angular ordering describes the measurements significantly better than the conventional one. 

It is important to note, that there are no free parameters left: once the TMD parton density is determined, the initial state parton shower follows exactly the TMD parton distribution. The TMD parton distribution is the essential ingredient in the present calculation, and a precise
determination of the TMD parton distribution over a wide range in $x$, \kt\ and scale $\mu$ is an important topic. First steps towards a precision determination of the TMD densities from HERA measurements have been performed in Ref. \cite{Hautmann:2017fcj,Hautmann:2017xtx}.

\section{Conclusion}

A new calculation using off-shell matrix elements with TMD parton densities  supplemented with a newly developed initial state TMD parton shower has been presented. The calculation is based on the \katie\ package for an automated calculation of the partonic process in high-energy factorization, making use of TMD parton densities implemented in TMDlib. The partonic events are stored in an LHE file, similar to the conventional LHE files, but now containing the transverse momenta of the initial partons. The LHE files are read in by the \cascade\  package for the full TMD parton shower where events in HEPMC format are produced for further processing, like with Rivet.

We have determined a full set of TMD parton densities using the KMRW approach, which include all flavours and are valid over a wide range in $x$, \kt , and $\mu$. These TMD parton densities are available in TMDlib. 

We have developed an initial state TMD parton shower, including all flavors and following the TMD distribution, without the need for adjusting further parameters.

As an example of application we have calculated the azimuthal de-correlation of high \pt\ dijets as measured at the LHC and found very good agreement with the measurement. It is remarkable, that using TMDs with off-shell matrix element calculations covers already a larger phase space than is accessible in collinear higher order calculations. Including initial state TMD parton showers together with conventional final state parton showers gives a remarkably good description of the measurements, which opens the floor for a rich phenomenology at the LHC making use of the advantages of automatic off-shell matrix element calculations with a fully TMD consistent parton shower.

\section*{Acknowledgements}
HJU thanks the Foundation for Polish Science (FNP) for support with an Alexander von Humboldt Polish Honorary Research Scholarship, allowing extensive stays in Cracow where the present work was completed.
MB, KK, MS acknowledge the support of NCN grant DEC-2013/10/E/ST2/00656. 
AvH was supported by grant of National Science Center, Poland, No.\ 2015/17/B/ST2/01838.
MS is also partially supported by the  Israeli Science Foundation through grant 1635/16, 
by the BSF grants 2012124 and 2014707, by the COST Action CA15213 THOR
and by a Kreitman fellowship from the Ben Gurion University of the Negev.

\bibliographystyle{mybibstyle} 
\raggedright 

\bibliography{/Users/jung/Bib/hannes-bib}

\end{document}